\documentclass[journal]{IEEEtran}
\ifCLASSINFOpdf
\else
 \usepackage[dvips]{graphicx}
\fi

\usepackage{epsfig}
\usepackage{amssymb}
\usepackage{graphicx}
\usepackage{bm}
\usepackage{subfigure}
\usepackage{cite}
\usepackage{amsmath}
\usepackage{amsfonts}
\newcommand {\R} {\mathbb{R}}
\newcommand {\al} {\alpha}
\newcommand {\be} {\beta}
\newcommand {\ep} {\varepsilon}
\newcommand {\la} {\lambda}

\newcommand {\De} {\Delta}

\newcommand {\si} {\sigma}
\newcommand {\bu} {\bullet}
\newcommand {\ab} [1] {\langle{#1}\rangle}

\newcommand {\LT} {\mathrm{LT}}

\graphicspath{{figs/}{../ieee_graphics/}}

\begin{document}


\title{Connectivity of Random 1-Dimensional Networks}

\author{
 V. Kurlin and L.~Mihaylova 

\thanks{\emph{Corresponding author}: V.~Kurlin,
 vitaliy.kurlin@durham.ac.uk,
 Department Mathematical Sciences,
 Durham University,
 Durham DH1 3LE, UK }

\thanks{L.~Mihaylova,
 mila.mihaylova@lancaster.ac.uk,
 Department of Communication Systems,
 Lancaster University,
 Lancaster LA1 4WA, UK }  }

\maketitle

\begin{abstract}
An important problem in wireless sensor networks
 is to find the minimal number of randomly deployed sensors
 making a network connected with a given probability.
In practice sensors are often deployed one by one along a trajectory of a vehicle,
 so it is natural to assume that arbitrary probability density functions
 of distances between successive sensors in a segment are given.
The paper computes the probability of connectivity and coverage of 1-dimensional networks
 and gives estimates for a minimal number of sensors for important distributions.
\end{abstract}


\IEEEpeerreviewmaketitle


\section{Introduction}


Recently the problems of connectivity and coverage in wireless
 sensor networks have been extensively investigated \cite{akyildiz:2002:surveynetworks}.
One-dimensional networks are theoretically simple, but can be used
 in many practical problems such as monitoring of
 roads, rivers, coasts and boundaries of restricted areas.
Networks distributed along straight paths can provide nearly
 the same information about moving objects as 2-dimensional networks,
 but require less sensors and have a lower cost.
\smallskip

We derive the probability of connectivity of a 1-dimensional network containing
 \emph{finitely many} sensors deployed according to \emph{arbitrary} densities
 in contrast to \cite{desaiandmanjunath:2002:connectivity}.
We found an exact formula in the general case and explicit estimates
 for a minimal number of sensors for classical distributions.
The main novelty is the universal approach to computing the probability of connectivity,
 which leads to closed expressions for piecewise constant densities
 approximating an arbitrary density.
The feasibility of the proposed approach is demonstrated over different scenarios.
We deal with densities of \emph{distances} between successive sensors,
 not with the distributions of sensors themselves, because sensors of
 1-dimensional networks are often deployed one by one along a trajectory of a vehicle.
\smallskip

%
%
%
%


Suppose that a sink node at the origin $x_0=0$
 collects some information from other sensors.
Let $L$ be the \emph{length} of a segment, where $n$ sensors
 having a \emph{transmission radius} $R$ are deployed.
The sensor positions are supposed to be in increasing order, i.e.
 $0=x_0\leq x_1\leq\dots\leq x_n\leq L$.
Let $f_i(s)$ be the probability density function of the $i$-th distance $y_i=x_{i}-x_{i-1}$.
The probability that $y_{i}\in[0,l]$
 can be computed as $P(0\leq y_i\leq l)=\int\limits_0^l f_i(s)ds$.
The resulting network is \emph{connected} if the distance $y_i$
 between any successive sensors, including the sink node, is not greater than $R$.
\smallskip

We assume that the distances are independently distributed.
The densities $f_i$ depend on the practical way to deploy sensors.
We consider the transmission radius $R$ as an input parameter,
 because the range of available radii is often restrictive,
 while the number of sensors can be easily controlled in practice.
\medskip

\noindent
{\bf The Connectivity Problem.}
Find the minimal number of randomly deployed sensors in $[0,L]$
 such that the resulting network is connected with a given probability.
\medskip

\noindent
{\bf The Coverage Problem.}
Find the minimal number of randomly deployed sensors such that the network
 is connected and covers the segment $[0,L]$ with a given probability.
\medskip

The example below shows that connectivity of networks
 in dimensions 1 and 2 are closely related.
Distributing sensors from a vehicle along a path in a forest can result
 in a network located in a narrow road of some width $W$, see Fig.~1.
Assuming that $W<R$ and denoting the 2-dimensional positions of the sensors by
 $(x_1,z_1),\dots,(x_n,z_n)$, where the $n$ sensors are ordered by their $x$-coordinates,
  the coordinate $z_i\in[-W/2,W/2]$  can be represented as a deviation of
 the $i$-th sensor from the central horizontal segment $[0,L]$.

\begin{figure}[!h]
\centering
\includegraphics[width=2.5in]{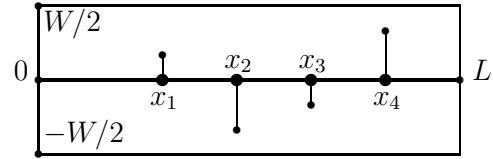}
\caption{A 2-dimensional network distributed in a narrow road}
\label{Fig1}
\end{figure}

If the 2-dimensional network is connected, i.e. each distance is not
 greater than the transmission radius $R$, then the Pythagoras theorem
 implies that $x_i-x_{i-1}\leq\sqrt{R^2-W^2}$ since $|z_i-z_{i-1}|\leq W$.
If the 1-dimensional network of the sensors $x_1,\dots,x_n$ projected to
 the horizontal segment $[0,L]$ is connected for the new transmission radius $\sqrt{R^2-W^2}$,
 then the original 2-dimensional network is also connected.
\smallskip

Similarly, if the 1-dimensional network of projections covers $[0,L]$,
 then the original 2-dimensional network covers the whole road $[0,L]\times[-W/2,W/2]$.
So if the width $W$ of the road can be assumed to be less than
 the original transmission radius $R$, then the connectivity and coverage problems
are reduced to the simpler problems for 1-dimensional networks.
\smallskip

The paper is organised as follows.
Related results on connectivity are reviewed in section~II.
In section~III we state the main theorems computing the probabilities of connectivity and coverage. Sections~IV, V, VI are devoted to explicit estimates of the minimal number of sensors for a uniform distribution, constant density with 2 parameters,
 truncated exponential and normal distribution.
Appendices~A--D contain proofs of the main theorems and corollaries
 including a method for computing the probability of connectivity for
 piecewise constant densities approximating any  density in practice.


\section{Related Results on Connectivity}
%
%
Many results on connectivity are \emph{asymptotic} in the number of sensors,
 see \cite{cheng:1898:critical,guptaandkumar:1998:critical} for 2-dimensional networks.
The network of $n$ sensors in the unit disk is connected with probability 1 if and only if
 the transmission radius $R$ is proportional to $\sqrt{(\ln n)/n}$ as $n\to\infty$
 \cite{panchapakesanandmanjunath:2001:transmission}, where $\ln$ means the logarithm to the base $e$.
These asymptotic results cannot be applied to real networks,
 because the rate of convergence is not clear.
\smallskip


The standard assumption for finite networks
 is the uniform distribution of sensors.
The authors of \cite{gupta:2006:topological} suppose that
 sensors are \emph{exponentially} distributed in a segment.
Papers \cite{panchapakesanandmanjunath:2001:transmission}
 and \cite{karamchandaniandmanjunathandiyer:2005:clustering}
 consider sensors having the \emph{Poisson}
 and \emph{exponential} distribution in square $[0,1]^2$, respectively,
 see also \cite{wan:2006:coveragerandom}, \cite{rametal:2007:path}.
\smallskip

An explicit analytical result on connectivity of finite networks
 was obtained in \cite{desaiandmanjunath:2002:connectivity},
 where $n$ sensors are uniformly distributed in $[0,L]$.
In this case the probability $P'_n$ of connectivity of the network
 was computed assuming $\binom{n-1}{i}=0 \mbox{ for }i\geq n$
$$P'_n=\sum\limits_{i=0}^{i<L/R}(-1)^i\binom{n-1}{i}(1-iR/L)^n.$$
The upper bound $i<L/R$ implies that $1-iR/L>0$, but
 the alternating inequality $P'_n\geq 0$ is still highly non-trivial
 and can hardly be proved by combinatorial methods.
This approach was generalised to the exponential distribution \cite{iyer:2006:connectivity1Dnetworks}.
\medskip

\begin{figure}[!h]
\centering
\includegraphics[width=2.5in]{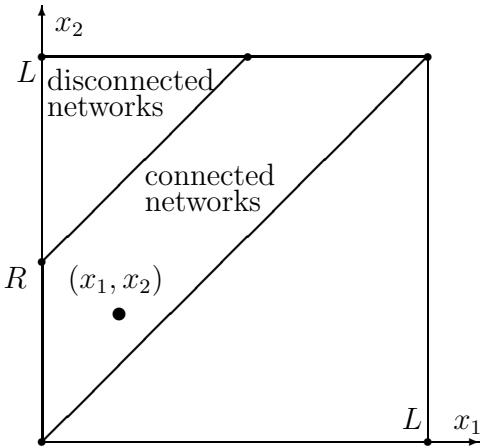}
\vspace{-3mm} \caption{The probability of connectivity for 2
uniformly distributed sensors} \label{Fig2}
\end{figure}

By the formula above for $n=2$ sensors having a transmission radius $R$,
 the probability of connectivity is $P'_2=1-(1-R/L)^2=2(R/L)-(R/L)^2$.
This is illustrated in Fig.~2, where a network of 2 sensors at $x_1,x_2$
 is represented by a point in the triangle $\{0\leq x_1\leq x_2\leq L\}$.
Then $P'_2$ is the area of the domain $\{0\leq x_2-x_1\leq R\}\cap[0,L]^2$ of
 connected networks divided by the area $L^2/2$ of the triangle.


\section{New Theoretic Results}



Recall that one deploys $n$ sensors having a transmission radius $R$
 in $[0,L]$ in such a way that the $i$-th distance $x_i-x_{i-1}$ between
 successive sensors has a probability density function $f_i(s)$ for $i=1,\dots,n$.
Assume that the densities $f_1,\dots,f_n$ are integrable and
 $\int\limits_{0}^{L} f_i(s)ds=1$, $i=1,\dots,n$.
Hence the $i$-th distance can take values from 0 to $L$.
So the $n$-th sensor may not be within $[0,L]$ and its position is bounded only by $nL$.
A network is \emph{proper} if all sensors are deployed in $[0,L]$.
In practice all networks are proper, because
 sensors are deployed along a fixed segment.
A proper network is \emph{connected} if the distance between
 any successive sensors, including the sink node at 0,
 is not greater than $R$, see Fig.~2.
\smallskip

We will compute the \emph{conditional} probability that a proper network
 is connected, i.e. the probability that the network
 is connected assuming that it is proper.
So the answer will be a fraction, the probability that the network
 is proper and connected over the probability that the network is proper.
The numerator and denominator will be evaluations
 of the function $v_n(r,l)$ defined recursively for $n\geq 0$ as follows:
\smallskip

\begin{tabular}{ll}
$v_0(r,l)=1$ & if $r,l>0$;\\
$v_n(r,l)=0$ & if $r\leq 0$ or $l\leq 0$;\\
$v_n(r,l)=1$ & if $r\geq l>0$, $n>0$;\\
$v_n(r,l)=\int\limits_0^r f_n(s)v_{n-1}(r,l-s)ds$ & if $r<l$, $n>0$.
\end{tabular}
\medskip

\noindent
{\bf The Probability Proposition.}
For $0<r\leq l$ in the above notations, $v_n(r,l)$ is the probability that an array of
 random distances $(y_1,\dots,y_n)$ with densities $f_1,\dots,f_n$, respectively,
 satisfies $\sum\limits_{i=1}^n y_i\leq l$ and $0\leq y_i\leq r$ for $i=1,\dots,n$.
\medskip

The variables $r,l$ play the roles of the upper bounds for the distance
 between successive sensors and the sum of distances, respectively.
Clearly $v_n(L,L)$ is the probability that a network is \emph{proper}, i.e. all sensors are
 in $[0,L]$, and $v_n(R,L)$ is the probability that a network is proper and connected.
\medskip

\noindent
{\bf The Connectivity Theorem.}
Let $n$ sensors $x_1,\dots,x_n$ having a transmission radius $R$ be deployed in $[0,L]$
 so that a sink node is fixed at $x_0=0$ and the distances $y_i=x_i-x_{i-1}$, $i=1,\dots,n$,
 have given probability density functions $f_1,\dots,f_n$.
Then the probability of connectivity of the resulting network is
 $P_n=\dfrac{v_n(R,L)}{v_n(L,L)}$, which is independent of the order of sensors,
 the function $v_n(r,l)$ was recursively defined above.
\medskip

Given a probability $p$, the answer to
 the Connectivity Problem from section I
 is the minimal number $n$ such that $P_n\geq p$.
A network of a sink node at $0$ and 1 sensor with at $y_1\in[0,L]$
 is connected with probability $P_1=P(0\leq y_1\leq R)=$\\
$=v_1(R,L)=\int\limits_0^R f_1(l)dl,\mbox{ since }v_1(L,L)=\int\limits_0^L f_1(l)dl=1.$
\medskip

\noindent
{\bf The Coverage Theorem.}
Under the conditions of the Connectivity Theorem,
 the probability that the network is connected and covers
 the segment $[0,L]$ is $\dfrac{v_n(R,L)-v_n(R,L-R)}{v_n(L,L)}$.
\medskip

The Connectivity Theorem leads to closed expressions for probability of connectivity
 and explicit estimates on a minimal number of sensors making a network connected
 with a given probability for classical densities in sections IV--VI.
The Connectivity and Coverage Theorems are proved in Appendix~A by generalising
 the analytical method from \cite{desaiandmanjunath:2002:connectivity}.


\section{The Uniform Distribution}

In this section we consider the simplest constant density $f(l)=1/L$ on $[0,L]$,
 i.e. the distances between successive sensors are uniformly distributed in $[0,L]$.
The formula for $P_n^u$ in the Uniform Corollary below can be compared with the formula
 for $P'_n$ from section~II obtained in \cite{desaiandmanjunath:2002:connectivity}
 for networks whose sensors (not distances) are uniformly distributed in $[0,L]$.
In the latter case there is no sink node at 0, see differences in Fig.~2--3.
In Fig.~2 the network of 2 sensors is represented by their positions $(x_1,x_2)$,
 while in Fig.~3 the same network is encoded by the distances $(y_1,y_2)=(x_1-0,x_2-x_1)$.
\medskip

\noindent
{\bf The Uniform Corollary.}
Under the conditions of the Connectivity Theorem,
 if the distances between successive sensors are uniformly distributed in $[0,L]$, then
 the probability of connectivity is $P_n^{u}=\sum\limits_{i=0}^{i<L/R}(-1)^i\binom{n}{i}(1-iR/L)^n$.
Set $Q=\dfrac{L}{R}-1$.
The network is connected with a given probability $p>2/3$ if
$$n\geq \dfrac{1}{2}\left(3(1-Q)+\sqrt{(3Q-1)^2+24Q^2\left(\dfrac{Q}{1-p}-1\right)}\;\right).$$
\medskip

\begin{figure}[!h]
\centering
\includegraphics[width=2.5in]{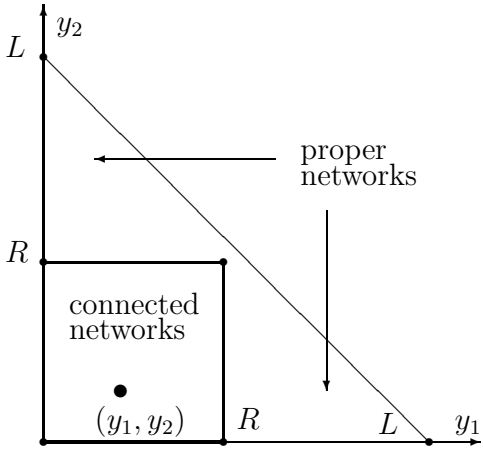}
\caption{The probability of connectivity for a sink node at 0 and 2
sensors
 with uniformly distributed distances $y_1=x_1-0$ and $y_2=x_2-x_1$ in $[0,L]$}
\label{Fig3}
\end{figure}

For $n=1$ the Uniform Corollary gives $P_1^u=R/L$, namely
 a network of a sink node at 0 and another $n=1$ sensor at a distance $y_1=x_1-0$
 is connected if and only if $y_1\leq R$, i.e. with probability $P_1^u=R/L$.
For $n=2$ one gets:
$$P_2^u=\left\{ \begin{array}{ll}
 2(R/L)^2 & \mbox{ if } R\leq L/2,\\
 4(R/L)-2(R/L)^2-1 & \mbox{ if } R\geq L/2.
 \end{array} \right.$$
If $R\leq L/2$, then the probability is the area of the square
 $\{0\leq y_1\leq R,\; 0\leq y_2\leq R\}$ divided by
 the area of the triangle $\{0\leq y_1,\; 0\leq y_2,\; 0\leq y_1+y_2\leq L\}$, see Fig.~3.
The lower bound in the Uniform Corollary is positive if $L\geq 2R$, because
 the 2nd term under the square root is non-negative for $p\in(0,1)$
 and the square root is not less than $3Q-1$.
\smallskip

The computational complexity of $P_n$ is linear in the number $n$ of sensors.
By the computational \emph{complexity} we mean the number of standard operations
 like multiplications and evaluating simple functions like $\ln(x)$.
A linear algorithm computing $P_n$ above initialises the array
 consisting of $n+1$ elements $L-iR$, $i=0,\dots,n$, then finds
 $\ln(L-iR)$, $n\ln(L-iR)$ and $\exp(n\ln(L-iR))=(L-iR)^n$.
The array of binomial coefficients $\binom{n}{i}$
 has $n+1$ elements and can be computed in advance.
So the total complexity of computing the probability $P_n^u$
 in the Uniform Corollary is $O(n)$.
\smallskip

Consider the segment of length $L=1$km and $n\leq 200$ sensors
 having transmission radius $R=50$m.
Suppose that a sink node is fixed at 0 and
 the distances between successive sensors
 have the same uniform distribution on $[0,L]$.
The graph in Fig.~4 shows the probability $P_n^u$ of connectivity
 computed in the Uniform Corollary.
The number $n$ of sensors varies from 1 to 200 on the horizontal axis.
\smallskip

\begin{figure}[!h]
\centering
\includegraphics[width=3.5in]{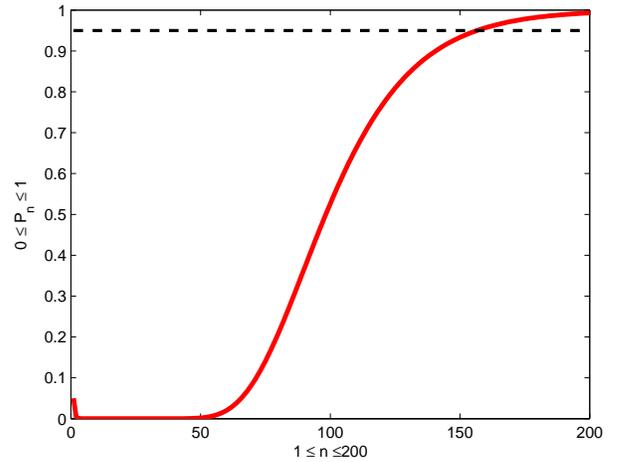}
\vspace{-5mm} \caption{The probability of connectivity for the
uniform distribution} \label{Fig4}
\end{figure}

The graph in Fig.~4 implies that after a certain value of $n$
 the probability $P_n^u$ of connectivity increases with respect to the number of sensors.
To solve the Connectivity Problem from section~I for a given probability $p$,
 we compute $P_n^u$ for all values from 1 to a minimum $n$ such that $P_n^u\geq p$.
Another method uses the estimate from the Uniform Corollary,
 which may not be optimal, but requires much less computations.
The exact minimal numbers and their estimates are in Table~1, where
 the network in $[0,L]$ with $L=1$km is connected with probability $p=0.95$.
For example, the minimal number of sensors for $R=50$m is 157, while the estimate is 905.
\medskip


\hspace*{-6mm}
\begin{tabular}{|l|c|c|c|c|c|}
\multicolumn{5}{c}{\textbf{Table 1. Simulations for the uniform distribution}} \\
\hline

Transmission Radius, m. & 200 & 100 & 50 & 25 & 10 \\
\hline

Min Number of Sensors & 29 & 69 & 157 & 349 & 982\\
\hline

Estimate of Min Number & 83 & 283 & 905 & 2610 & 10640\\
\hline
\end{tabular}
\medskip

Table~1 implies that the uniform distribution is
 very idealised and can not be useful in practice.
If one deploys sensors of transmission radius $R=50$m
 non-randomly at regular intervals 49m, than 21 sensors are
 enough to make the network connected and cover $[0,L]$.
The estimate from the Uniform Corollary is too rough, because of
 the term $24Q^3/(1-p)$ under the square root, which can be very large.
The uniform distribution is extended to a more practical case in section~V.


\section{A constant density with 2 parameters}

In this section we consider the constant density $f(l)=1/(b-a)$ over any segment
 $[a,b]\subset[0,L]$, which generalises the uniform distribution from section IV.
Practically the distribution means that each sensor is thrown at a distance
 uniformly varying between $a$ and $b$ from the previously deployed sensor.

\begin{figure}[!h]
\centering
\includegraphics[width=2.5in]{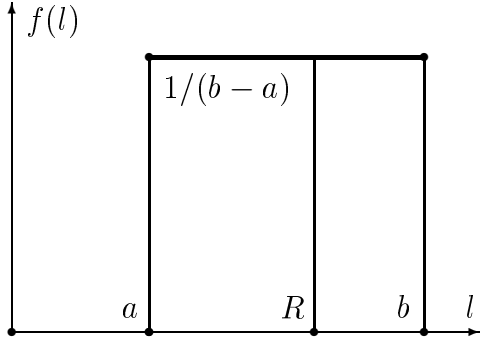}
\vspace{-3mm} \caption{The constant density over the segment
$[a,b]$} \label{Fig5}
\end{figure}

The left endpoint $a$ should be less than the transmission radius $R$,
 otherwise no sensor communicates with its neighbours.
The mathematical \emph{expectation} of the distance between successive sensors
 is $(a+b)/2$, while $(b-a)^2/12$ is the \emph{variance} of the distance.
For example, for a network of a sink node at 0 and 1 sensor at $y_1$,
 the probability of connectivity is $P(0\leq y_1\leq R)=(R-a)/(b-a)$, see Fig.~5.
\medskip

\noindent
{\bf The Constant Corollary.}
If in the Connectivity Theorem the distances between successive sensors
 have the density $f(l)=1/(b-a)$ on $[a,b]$, then the probability of connectivity is
$$P_n^c=\dfrac{ \sum\limits_{k=0}^n (-1)^k \binom{n}{k} (L-a(n-k)-Rk)^n }{
  \sum\limits_{k=0}^n (-1)^k  \binom{n}{k} (L-a(n-k)-bk)^n }.$$
The network is connected with a given probability $p$ if \\
 $n\geq\max\left\{ \dfrac{3}{2}+\sqrt{\dfrac{1+5p}{1-p}},\; 1+\dfrac{L-b}{a} \right\}$
 and $\dfrac{a+b}{2}\leq R\leq b$.
\medskip

The sums include all expressions taken to the power $n$ if they are positive.
The complexity to compute $P_n^c$ is $O(n)$.
Each of the terms in both sums requires $O(1)$ operations similarly to the Uniform Corollary.
For $n=1$ one gets $P_1^c=\dfrac{(L-a)-(L-R)}{(L-a)-(L-b)}=\dfrac{R-a}{b-a}$ as expected above.
If the given probability $p$ is too close to 1 then the estimate
 from the Constant Corollary depends on $p$, e.g. $n\geq 776$ for $p=0.9999$,
 but in all reasonable cases the maximum is achieved at
 the second expression $1+(L-b)/a$ independent of $p$.
The restrictions $\dfrac{a+b}{2}\leq R\leq b$ seem to be natural
 saying that the distance between successive sensors is likely to be
 less than $R$ since $[0,R]$ covers more than a half of $[a,b]$.
\smallskip

Each distance $x_i-x_{i-1}$ between successive sensors belongs to $[a,b]$.
Such a network lies within $[0,L]$ only if $an\leq L$,
 hence the number of sensors should satisfy $n<L/a$.
In the boundary case $an=L$ all sensors should be located at the exact positions
 $x_i=ia/n$, $i=1,\dots,n$, which clearly happens with probability 0,
 so the numerator vanishes for $L=an$ in the Constant Corollary.
If $n>L/a$ then $n$ sensors can not be within $[0,L]$ according to the density $1/(b-a)$.
\smallskip


\begin{figure}[!h]
\centering
\includegraphics[width=3.5in]{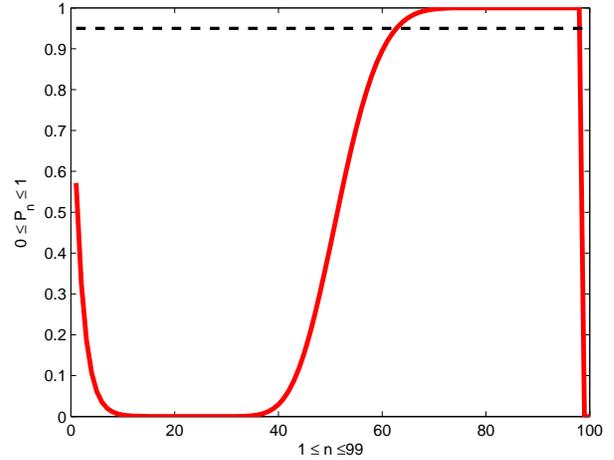}
\vspace{-3mm} \caption{The probability of connectivity, $R=50$m,
$a=0.2R$, $b=1.6R$} \label{Fig6}
\end{figure}

\noindent
{\bf Table 2.}
The case of the constant density over $[0.2R,1.6R]$.
\smallskip

\hspace*{-5mm}
\begin{tabular}{|l|c|c|c|c|c|}
\hline
Transmission Radius, m. & 200 & 150 & 100 & 50 & 25\\
\hline

Min Number of Sensors & 14 & 19 & 30 & 63 & 132\\
\hline

Estimate of Min Number & 18 & 27 & 43 & 93 & 193 \\
\hline

Max Number of Sensors & 25 & 34 & 50 & 100 & 200\\ \hline
\end{tabular}\\


Figs.~6--8 show the probability of connectivity for
 different segments $[a,b]$ depending on the radius $R=50$m.
The graph in Fig.~6 is the probability $P_n^c$ of connectivity
 for $1\leq n\leq 100$, $L=1$ km, $R=50$m, $a=0.2R$, $b=1.6R$.
If the required probability of connectivity is $p=0.95$ and
 the transmission radius is 50m, then the minimal number of sensors is 63.
\smallskip

The maximal possible number of sensors is $L/a=100$, i.e. $P_{100}^c=0$ since
 the sensors should be fixed at exact positions in $[0,L]$, which explains the drop to 0 in Fig.~6.
The minimal number of sensors decreases when the length $b-a$ decreases.
The maximum number of sensors in Table 2 is $L/a$,
 which gives probability 0 in this extreme case.
All numbers slightly less than the maximum give a probability close to 1.
More exactly we may subtract $b/a-1=7$, see Table~2, which follows from
 the second restriction $n\geq 1+(L-b)/a$.
\medskip


\begin{figure}[!h]
\centering
\includegraphics[width=3.5in]{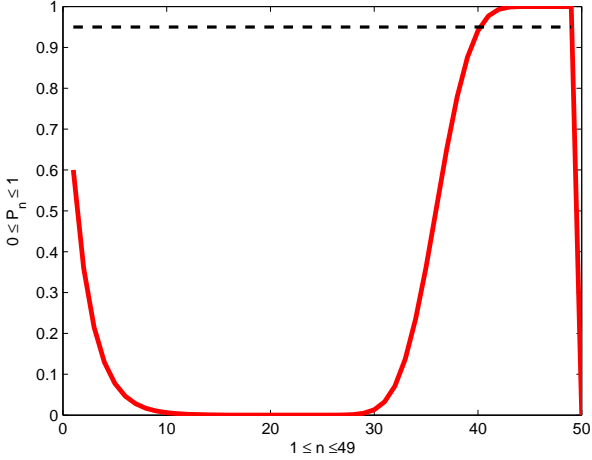}
\caption{The probability of connectivity, $R=50$m, $a=0.4R$, $b=1.4R$}
\label{Fig7}
\end{figure}

\noindent
{\bf Table 3.}
The case of the constant density over $[0.4R,1.4R]$.
\medskip

\hspace*{-5mm}
\begin{tabular}{|l|c|c|c|c|c|}
\hline Transmission Radius, m. & 200 & 150 & 100 & 50 & 25\\
\hline

Min Number of Sensors & 10 & 13 & 20 & 41 & 83 \\
\hline

Estimate of Min Number & 10 & 14 & 22 & 47 & 97 \\
\hline

Max Number of sensors & 13 & 17 & 25 & 50 & 100\\
\hline
\end{tabular}
\medskip


\begin{figure}[!h]
\centering
\includegraphics[width=3.5in]{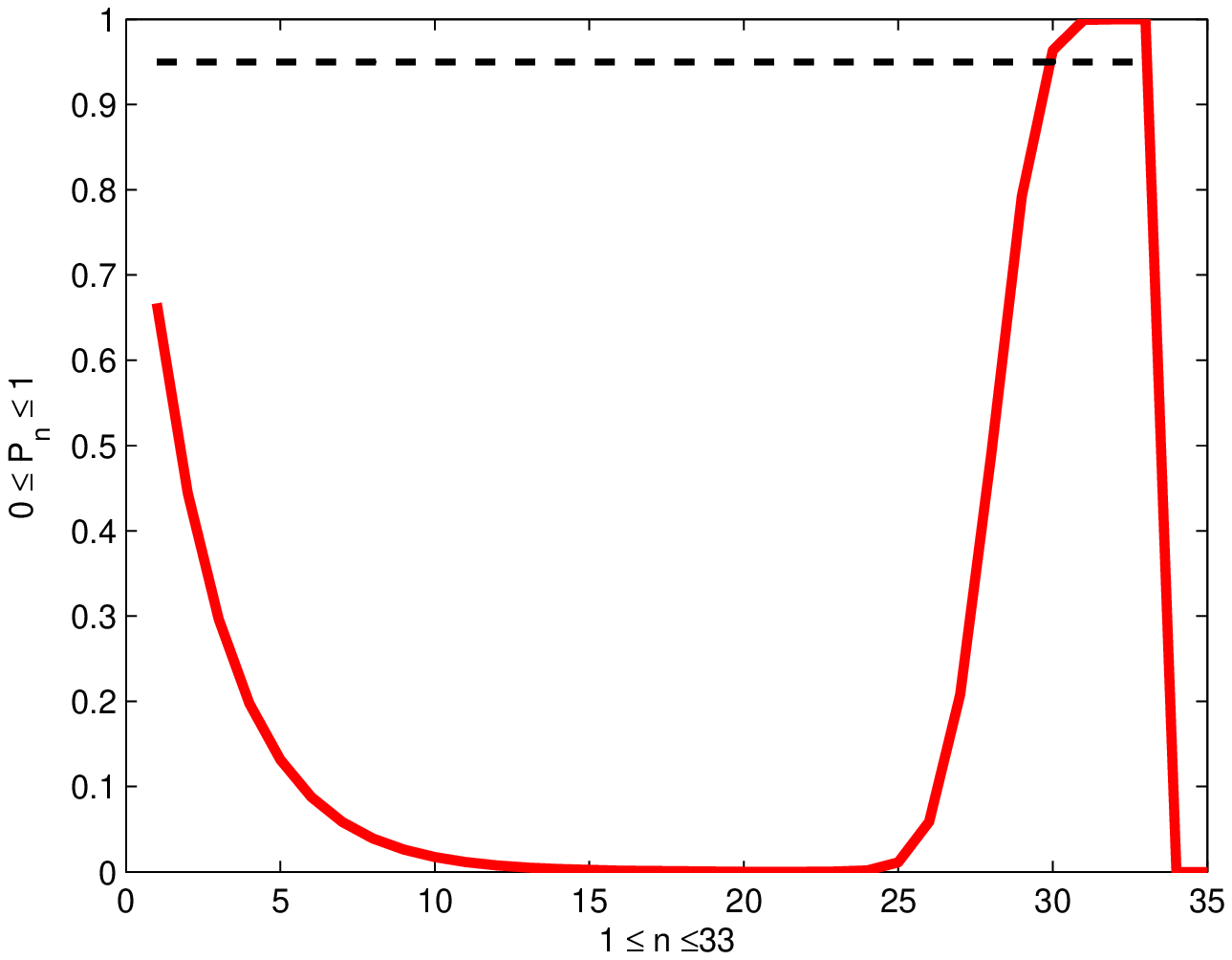}
\vspace{-3mm} \caption{The probability of connectivity, $R=50$m,
$a=0.6R$, $\rho=1.2R$} \label{Fig8}
\end{figure}

\noindent
{\bf Table 4.}
The case of the constant density over $[0.6R,1.2R]$.
\medskip

\hspace*{-5mm} \begin{tabular}{|l|c|c|c|c|c|}
 \hline Transmission Radius, m. & 200 & 150 & 100 & 50 & 25\\
\hline

Min Number of Sensors & 8 & 10 & 15 & 31 & 61 \\
\hline

Estimate of Min Number & 8 & 11 & 15 & 32 & 65 \\
\hline

Estimate of Max Number & 10 & 13 & 17 & 34 & 67\\
\hline
\end{tabular}
\medskip

Tables~2--4 imply that the required number of sensors making
 a network connected decreases if the ratio $(b-a)/R$ decreases.
For $b-a\leq R$ the estimate from the Constant Corollary is very close
 to the exact minimal number of sensors when  sensors
 are deployed non-randomly at a distance slightly less than $R$.
So the found estimate for the minimal number of sensors
 requires few computations and can be useful.


\section{Exponential and Normal Distributions}

Here we state partial results for 2 other classical distributions.
For the exponential density over $[0,L]$, we compute the exact probability of connectivity, but
 the simplest estimate for the minimal number of sensors is the same as for the uniform distribution.
For the normal density, it is hard to compute the probability of connectivity explicitly,
 but a reliable estimate for a maximal number of sensors will be derived.
\smallskip

Consider the exponential distribution $f(s)=\la e^{-\la s}$, $\la>0$.
It is used for modelling the wait-time until the next event in a queue.
Since sensors are deployed in $[0,L]$, we consider
 the truncated density $f(s)=c e^{-\la s}$
 on $[0,L]$ and $f(s)=0$ otherwise.
The condition $\int\limits_0^L f(s)ds=1$
 gives $c=\dfrac{\la}{1-e^{-\la L}}$.
\medskip

\noindent
{\bf The Exponential Corollary.}
If in the Connectivity Theorem the distances between successive sensors have
 the exponential density $f(s)=c e^{-\la s}$ in $[0,L]$, then the probability
 of connectivity is $P_n^e=\dfrac{v_n(R,L)}{v_n(L,L)}$, where $v_n(r,l)=$
$$=\sum\limits_{i=0}^{i<l/r} (-1)^i\binom{n}{i} \dfrac{e^{-i\la r}}{\lambda^n}
 \left(1- e^{-\lambda(l-ir)} \sum\limits_{j=0}^{n-1} \dfrac{\la^j(l-ir)^j}{j!} \right).$$
\medskip

The estimate for a minimal number of sensors from the Uniform Corollary
 holds in this case, which can be proved analytically, but easily follows
 from the fact that the exponential density monotonically decreases on $[0,L]$,
 hence the distance between successive sensors will be smaller on average
 than for the constant density over $[0,L]$, i.e. the network is more likely to be connected.
The computational complexity of $P_n^e$ in Corollary~2 is $O(n^2)$,
 because each expression in the brackets requires $O(n)$ operations
 as in Corollary~1 assuming that $\ln(x)$ and $\exp(x)$
 can be computed in $O(1)$ operations.
\smallskip

The Exponential Corollary implies that $v_n(L,L)=1-e^{-\la L}\sum\limits_{j=0}^{n-1} (\la L)^j/j!$
Indeed the term corresponding to $i=1$ vanishes if $r=l$.
The sum $\sum\limits_{j=0}^{n-1} \dfrac{\la^j(l-ir)^j}{j!}$
 converges rapidly to $e^{\la(l-ir)}$ as $n\to\infty$.
Hence the expression in the brackets from the Exponential Corollary
 is very close to 0 even for small $n$.
Then $P_n^e$ is a ratio of tiny positive values of order $10^{-10}$ or less.
The computation of $P_n^e$ very fast accumulates
 a big arithmetic error even for small $n$.
The exponential decreasing of $ce^{-\la s}$ means that
 the sensors are distributed very close to each other
 and cover $[0,L]$ with little probability.
So the exponential distribution seems to be rather unpractical
 for modelling distances between successive sensors.
\smallskip

Finally we consider the remaining classical distribution, the truncated normal density over $[0,L]$,
 i.e. $f(s)=\dfrac{c}{\si\sqrt{2\pi}}e^{-(s-\mu)^2/2\si^2}$, where
 the constant $c$ guarantees that $\int_0^L f(s)ds=1$.
The normal density has exponentially decreasing tails, so
 distances between successive sensors are likely to be close to $\mu$.
Hence the mean $\mu$ should be less than the transmission radius $R$
 and the number of sensors $n$ can not be greater than $L/\mu$,
 otherwise last sensors are likely to be outside $[0,L]$.
That is why the Normal Corollary below gives an upper bound for the number of sensors
 making a network connected, not a lower bound as in previous corollaries.
\medskip

\noindent
{\bf The Normal Corollary.}
If in the Connectivity Theorem the distances between successive sensors
 have the truncated normal distribution on $[0,L]$ with
 a mean $\mu$ and standard deviation $\si$ then
 the network is connected with a given probability $p$ for
$$n\leq\min\left\{\dfrac{p(1-p)}{\ep},\; \dfrac{(\sqrt{4\mu L+\si^2\Phi^{-2}(p)}-\si\Phi^{-1}(p))^2}{4\mu^2}\right\},$$
$$\Phi(x)=\dfrac{1}{\sqrt{2\pi}} \int\limits_{-\infty}^x e^{-s^2/2}ds,\;
  \ep=\Phi\left(-\dfrac{\mu}{\si}\right)+1-\Phi\left(\dfrac{R-\mu}{\si}\right).$$
\smallskip

The standard normal distribution $\Phi(x)$ is not elementary, but its values have been tabulated.
The table below shows estimates for the maximal number of sensors
 normally distributed in $[0,L]$ with $L=1$ km, $\mu=0.6R$, $\si=0.1R$ in such a way that
 the resulting network is connected with probability $p=0.9975$.
Then $\Phi^{-1}(p)\approx 2.8$, $\ep\approx 0.000063$ and
 the first upper bound in the Normal Corollary gives $n\leq p(1-p)/\ep\approx 40$,
 which is the overall upper bound for $R=25$m.
For radii $R\geq 50$m the second upper bound is smaller that the first one and
 is close to $L/\mu$, the exact number of sensors when all distances
 are not random and equal to $\mu$, because
 $\si\Phi^{-1}(p)/R\approx 0.28$ is rather small.
\medskip

\noindent {\bf Table 5.} The case of the normal density, $\mu=0.6R$,
$\si=0.1R$. 
\smallskip

\hspace*{-5mm}
\begin{tabular}{|l|c|c|c|c|c|}
\hline
Transmission Radius, m. & 200 & 150 & 100 & 50 & 25\\
\hline

Estimate of Max Number & 7 & 11 & 16 & 33 & 40\\
\hline
\end{tabular}
\medskip

The estimates from Table~5 are close to optimal, e.g. for the radius $R=150$m
 the non-random distribution of sensors at distance 149m apart requires 6 sensors
 not including the sink node at 0, while the estimate above gives 11.
The ratio $6/11$ is close to the mean $\mu/R=0.6$ since distances
 between successive sensors should be around the average $\mu=0.6R$.

\section{Conclusions}

We would like to emphasise that the main result of the paper is a new method of analytical
 computing the probability of connectivity of random 1-dimensional networks leading to
 explicit formulae for piecewise constant densities approximating an arbitrary density.
The found estimates for a minimal number of sensors making a network connected
 suggest that a constant and normal densities over a segment can be
 more economic than other other classical distributions.

\smallskip
Open issues for the future research are the following:

\noindent
\emph{i}) computing analytically the exact probability of
connectivity in the case when the distances between successive
sensors have
 a truncated normal distribution over $[0,L]$;
\smallskip

\noindent
\emph{ii}) finding an optimal distribution of distances between
successive sensors in $[0,L]$ for
 a given number of sensors to maximise the probabilities of connectivity and coverage;
\smallskip

\noindent \emph{iii})
 extending the suggested approach of sensor distributions
 to non-straight trajectories filling a 2-dimensional area.


\appendices

\section{Proofs of the Main Theorems}

First we recall the notion of the convolution and Laplace transform used in the proof
 of the Connectivity Theorem, Coverage Theorem and Corollaries from sections~IV-VI.
The \emph{convolution} of functions $f,g$ is
 $f*g(s)=\int\limits_{-\infty}^{+\infty}f(l)g(s-l)dl$.
The convolution is commutative, associative, distributive
 and respects constant factors, i.e.
 $(cf)*g=c(f*g)$, $f*g=g*f$, $(f*g)*h=f*(g*h)$, $f*(g+h)=f*g+f*h$.
The convolution plays a very important role in probability theory,
 because the probability density of the sum of 2 random variables
 is the convolution of the densities of the variables.
\smallskip

Given a function $f(l)$ and $r>0$, introduce the \emph{truncated} function
 $f^{[r]}(l)=f(l)$ for $l\in[0,r]$ and $f^{[r]}(l)=0$ otherwise.
Let $u(l)$ be the \emph{unit step} function
 equal to 1 for $l\geq 0$ and equal to 0 for $l<0$.
Then the truncated function $f^{[r]}(l)$ is $f(l)(u(l)-u(l-r))$.
Below we use the \emph{partial} convolution $f(r,l)*g(r,l)$
 considered only for the argument $l$, while $r$ remains constant.
The following lemma rephrases the recursive definition of $v_n(r,l)$
 in terms of convolutions.
\medskip

\noindent
{\bf Lemma~1.}
Given densities $f_1,\dots,f_n$, the function $v_n(r,l)$
 from Section~III is $f_n^{[r]} * \dots * f_1^{[r]} * u(l)$, $r<l$, $n>0$.
\medskip

\noindent
\emph{Proof of Lemma~1} is by induction on $n$.
The base $n=1$ is trivial:
 $f_1^{[r]}* u(l)=\int\limits_0^r f_1(s)u(s-l)ds=\int\limits_0^r f_1(s)ds=v_1(r,l)$
 since $s\leq r<l$.
The inductive step follows from the recursive definition of $v_n$ in section III:
 $v_n(r,l)=f_n^{[r]}* v_{n-1}(r,l)$.
\hfill $\square$
\medskip

The \emph{Laplace} transform of a function $f(l)$ is the function
 $\LT\{f(l)\}(s)=\int\limits_ {0}^{+\infty}e^{-sl}f(l)dl$.
The Laplace transform is a linear operator converting
 the convolution into the product, i.e.  $\LT\{af+bg\}=a\LT\{f\}+b\LT\{g\}$,
 $\LT\{f*g\}=\LT\{f\}\LT\{g\}$.
The inverse Laplace transform $\LT^{-1}$ is also a linear operator.
The following well-known properties of the Laplace transform
 can be easily checked by integration.
\medskip

\noindent
{\bf Lemma 2.}
For any $\al,\be$ and integer $m\geq 0$ one has\\
(a) $\LT\{l^m u(l)\}=\dfrac{m!}{s^{m+1}}$,\\
(b) $\LT\{e^{-\alpha l} l^m u(l) \}=\dfrac{m!}{(s+\alpha)^{m+1}}$,\\
(c) $\LT\{(l-\beta)^m u(l-\beta) \}=\dfrac{m!e^{-\beta s}}{s^{m+1}}$ and\\
(d) $\LT\{e^{-\alpha(l-\beta)} (l-\beta)^m u(l-\beta) \}=\dfrac{m!e^{-\beta s}}{(s+\alpha)^{m+1}}$.
\hfill $\square$
\medskip

Lemma~2 allows one to compute the inverse Laplace transform,
 e.g. Lemma~2(a) implies that $\LT^{-1}\{1/s\}=u(l)$.
Lemma~3 provides a powerful method for computing
 the function $v_n(r,l)$ used in the Connectivity Theorem.
\medskip

\noindent
{\bf Lemma~3.}
Given probability densities $f_1,\dots,f_n$ on $[0,L]$,
 set $g(s)=\LT\{f_n^{[r]}(l)\}\cdot\ldots\cdot\LT\{f_1^{[r]}(l)\}/s$.
Then the function $v_n(r,l)$ is
 the inverse Laplace transform $\LT^{-1}\{g(s)\}(l)$.
\medskip

\noindent
\emph{Proof of Lemma~3}.
One has $v_n(r,l)=f_n^{[r]} * \dots * f_1^{[r]}* u(l)$ by Lemma~1.
Set $g(s)=\LT\{v_n(r,l)\}$ and $g_i(s)=\LT\{f_i^{[r]}(l)\}$, $i=1,\dots,n$.
The Laplace transform is considered with respect to $l$,
 the variable $r$ is a fixed parameter.
The Laplace transform converts the convolution into the product,  hence $g(s)=g_1(s)\dots g_n(s)/s$
 as expected since $\LT\{u(l)\}=1/s$ By Lemma~2(a).
The order in the product $g(s)$ does not matter as the convolution is commutative.
So any reordering of the densities gives the same result and
 the probability of connectivity does not depend on this order.
\hfill $\square$
\medskip

\noindent
\emph{Proof of the Connectivity Theorem}.\\
Let $0=x_0\leq x_1\leq\cdots\leq x_n\leq L$ be the positions
 of a sink node and $n$ sensors.
Suppose that the distances $y_i=x_i-x_{i-1}$, $i=1,\ldots,n$,
 are independent and have probability densities $f_i(s)$.
Any network can be represented by ordered sensors $(x_1,\dots,x_n)$
 or, equivalently, by the distances $(y_1,\dots,y_n)$ between successive sensors.
Then the conditional probability of connectivity
 is the probability that the network is proper and connected, i.e.
 $\sum\limits_{i=1}^n y_i\leq L$ and $0\leq y_i\leq R$, divided by
 the probability that the network is proper, i.e.
 $\sum\limits_{i=1}^n y_i\leq L$ and $0\leq y_i\leq L$.
Hence the required formula $P_n=\dfrac{v_n(R,L)}{v_n(L,L)}$
 for the conditional probability of connectivity follows from
 the Probablity Proposition stated in section~III.
Permuting densities leads to the same probability $P_n$
 due to commutativity of the convolution from Lemma~1.
\hfill $\square$
\medskip

\noindent
\emph{Proof of the Probability Proposition}.\\
We illustrate the proof first in the partial cases $n=1,2$.
For $n=1$ and $L>r$, $P(0\leq y_1\leq r)=\int\limits_0^r f_1(s)ds=v_1(r,l)$
 and $P(0\leq y_1\leq l)=\int\limits_0^l f_1(s)ds=v_1(l,l)$ as expected.
\smallskip

For $n=2$, let the distance $y_2$ belong to
 $[s,s+\Delta]\subset[0,r]$ for some small $\Delta>0$.
The probability of this event $E$ is $P(E)=P(s\leq y_2\leq s+\De)\approx f_2(s)\Delta$,
 the area of the narrow rectangle below the graph of $f_2$ over $[s,s+\Delta]$.
The random variables $y_1=x_1-x_0$ and $y_2=x_2-x_1$ are assumed to be independent.
Then the probability of connectivity is
$$P(E)P(0\leq y_1\leq l-s)\approx f_2(s)\Delta\cdot v_1(r,l-s).$$
The total probability is the limit sum of the above quantities
 over the intervals $[s,s+\Delta]$ covering $[0,R]$ when $\Delta\to 0$.
Hence the probability is $\int\limits_0^r f_2(s) v_1(l-s)ds=v_2(r,l)$.
\smallskip

We will prove the general case $n>1$ by induction on $n$.
If the network is proper and connected then the $n$th distance
 $y_n=x_n-x_{n-1}\leq 0$ is not greater than $r$
 and not greater than $l-\sum\limits_{i=1}^{n-1}y_i$.
The former condition means that the last sensor is close enough to the previous one.
The latter condition guarantees that all the sensors are in the segment $[0,l]$.
\smallskip

Split $[0,r]$ into equal segments of a small length $\Delta$.
Suppose for a moment that $f_n(s)$ is constant on each segment
 $[s,s+\Delta]$, where $s=j\Delta$, $j=0,\dots,[r/\Delta]-1$.
The general case will be obtained by taking the limit under $\Delta\to 0$.
\smallskip

The probability $P(y_n\in[s,s+\Delta])$ is approximately $f_n(s)\Delta$, the area below
 the graph of $f_n(s)$ which is assumed to be constant over a short segment $[s,s+\Delta]$.
The probability that the $n-1$ sensors form a connected network in $[0,l-y_n]$
 is approximately $v_{n-1}(r,l-s)$ by the induction hypothesis.
\smallskip

Since the distances are distributed independently,
 the joint probability is $f_n(s)\Delta\cdot v_{n-1}(r,l-s)$.
The total probability is $v_n(r,l)$, the limit sum over all these events as $\Delta\to 0$:
$$\sum\limits_{j=1}^{[r/\Delta]} f_n(j\Delta)\Delta\cdot v_{n-1}(r,l-j\Delta)\to
 \int\limits_0^r f_n(s) v_{n-1}(r,l-s)ds.$$
The final expression above is the standard definition of
 the Riemann integral of $f_n(s)v_{n-1}(r,l-s)$ as a limit sum.
\hfill $\square$
\medskip

\noindent
\emph{Proof of the Coverage Theorem}.\\
By the Probability Proposition  $v_n(R,L)$ is the probability of the event $E(L)$ that
 $n$ sensors are deployed in $[0,L]$ and form a connected network.
The network covers $[0,L]$ if also
 at least one sensor is in $[L-R,L]$, i.e. $E(L-R)$ does not happen.
Hence the probability that the network is proper, connected
 and covers $[0,L]$ is $v_n(R,L)-v_n(R,L-R)$.
So the required conditional probability assuming that
 all sensors are in $[0,L]$ is equal to $\dfrac{v_n(R,L)-v_n(R,L-R)}{v_n(L,L)}$
 as required.
\hfill $\square$


\section{Proofs of the Main Corollaries}

The iterated convolutions respect constant factors, i.e.
$$(c_n f_n)^{[r]}* \dots *(c_1 f_1)^{[r]}*u=
  c_n\dots c_1 f_n^{[r]}*\dots * f_1^{[r]}*u.$$
Hence we may consider probability densities without extra factors if
 we are interested only in the conditional probability $P_n$
  from the Connectivity Theorem.
Indeed, the product of these factors will cancel dividing $v_n(R,L)$ by $v_n(L,L)$.
\medskip

\noindent
\emph{Proof of the Uniform Corollary}.\\
Let $f(l)$ be the uniform density over $[0,L]$, i.e.
 $f(l)=1/L$ on $[0,L]$, $f(l)=0$ otherwise.
Let $f^{(n*)}*g$ be the $n$-th iterated convolution,
 e.g. $f^{(2*)}*g=f*(f*g)$.
\smallskip

Lemma~3 gives a straightforward method to compute $v_n(r,l)=(f^{[r]})^{(n*)}*u$,
 where $u(l)$ is the unit step function, i.e.
 $u(l)=1$ for $l\geq 0$ and $u(l)=0$ for $l<0$.
\smallskip

Assume that $f^{[r]}(l)=u(l)-u(l-r)$ forgetting about front factors.
Lemma~2(c) implies that $\LT\{f^{[r]}\}=\dfrac{1-e^{-rs}}{s}.$
By Lemma~3 one has $v_n(r,l)=\LT^{-1}\{g(s)\}$, where
$$g(s)=\dfrac{(1-e^{-rs})^n}{s^{n+1}}=
 \sum\limits_{i=0}^n (-1)^i\binom{n}{i} \dfrac{e^{-irs}}{s^{n+1}}.$$
$$\mbox{By Lemma~2(b) }
 \LT^{-1}\left\{\dfrac{e^{-irs}}{s^{n+1}} \right\}(l)=\dfrac{(l-ir)^n}{n!} u(l-ir).$$
Replacing $u(l-ir)$ by the upper bound $i<l/r$, we get
$$v_n(r,l)=\LT^{-1}\{g(s)\}(l)
 =\sum\limits_{i=0}^{i<l/r} (-1)^i\binom{n}{i} \dfrac{(l-ir)^n}{n!}.$$

By the Connectivity Theorem the denominator of $P_n^u=v_n(R,L)/v_n(L,L)$ is $v_n(L,L)=L^n/n!$
Hence we may divide each term $(L-iR)^n/n!$ in $v_n(R,L)$ by $L^n/n!$,
 which gives the final formula from the Uniform Corollary.
\smallskip

Now we prove the estimate for a minimal number of sensors
 making the network connected with a given probability $p$,
  i.e. we should check that the probability $P_n^u\geq p$ if
 $$n\geq \dfrac{1}{2}\left(3(1-Q)+\sqrt{(3Q-1)^2+24Q^2\left(\dfrac{Q}{1-p}-1\right)}\;\right),$$
 where $Q=(L/R)-1$.
The idea is to simplify the inequality $P_n^u\geq p$ replacing $P_n^u$ by smaller and simpler expressions,
 which will lead to the required lower bound for $n$ above.
Setting $q=R/L$,  the probability from the Uniform Corollary
 becomes the alternating sum starting as follows:
$$P_n^u=1-n(1-q)^n+\binom{n}{2}(1-2q)^n-\binom{n}{3}(1-3q)^n+\cdots$$
The sum involves only positive terms of the form $1-iq$.
First we check that $P_n^u\geq 1-n(1-q)^n$ forgetting about the remaning terms.
It suffices to show that every odd term $\binom{n}{2k+1}(1-(2k+1)q)^n$ is not
 greater than the previous even one $\binom{n}{2k}(1-2kq)^n$ for $k\geq 1$.
The last inequality is equivalent to
 $\left(\dfrac{1-2kq}{1-(2k+1)q}\right)^n\geq \dfrac{n-2k}{2k+1}$.
Replace the left hand side by the smaller expression $(1-q)^{-n}$
 and the right hand side by the greater expression $n/3$ using $k\geq 1$.
\smallskip

The resulting inequality $(1-q)^{-n}\geq n/3$ is weaker than the simplified inequality
 $P_n^u\geq 1-n(1-q)^n\geq p$ for $p\in(2/3,1)$, i.e. $(1-q)^{-n}\geq n/(1-p)$.
We check that $(1-q)^{-n}\geq n/(1-p)$
 holds under the required restriction on $n$.
Since
$$Q=(L/R)-1=(1-q)/q \mbox{ then } (1-q)^{-n}=(1+1/Q)^n.$$
Therefore the proof finishes by the following lemma.
\medskip

\noindent
{\bf Lemma 4.}
The lower bound for $n$ from the Uniform Corollary
 implies that $(1+1/Q)^n\geq n/(1-p)$.
\medskip

\noindent
\emph{Proof of Lemma~4.}
Expand the Taylor series $(1+1/Q)^n=$
$$=1+n/Q+n(n-1)/2Q^2+n(n-1)(n-2)/6Q^3+\cdots$$
Leaving the terms of degrees 1,2,3 only makes the inequality
 $(1+1/Q)^n\geq n/(1-p)$ stronger, hence it suffices to prove
$$n/Q+n(n-1)/2Q^2+n(n-1)(n-2)/6Q^3\geq n/(1-p).$$
Multiplying boths sides by $6Q^3/n$, we get
$$6Q^2+3Q(n-1)+(n-1)(n-2)\geq 6Q^3/(1-p),$$
$$(n-1)^2 +(3Q-1)(n-1)+6Q^2(1-Q/(1-p))\geq 0.$$
The quadratic inequality holds if $n-1$ is not less than the 2nd
$$\mbox{ root }\dfrac{1}{2}\left(1-3Q+\sqrt{(3Q-1)^2+24Q^2\left(\dfrac{Q}{1-p}-1\right)}\;\right),$$
which is equivalent to the required condition on $n$.
\hfill $\square$
\medskip

\noindent
\emph{Proof of the Constant Corollary}. \\
The truncated constant density over $[a,b]$
 without extra factors is $f^{[r]}(l)=u(l-a(r))-u(l-b(r))$,
 where $[a(r),b(r)]=[a,b]\cap[0,r]$ is the domain, where
 the probability density is defined and restricted to $[0,r]$.
For instance, if $0<a<R<b<L$ then $a(L)=a(R)=a$ and $b(L)=b$, $b(R)=R$.
\smallskip

By the Connectivity Theorem and Lemma~1 the probability $P_n^c$
 is expressed in terms of $v_n(r,l)=(f^{[r]})^{(n*)}*u(l)$.
Lemma~2(c) for $m=0$, $\be=a(r)$, $\be=b(r)$ implies that
$$\LT\{ u(l-a(r))-u(l-b(r)) \}=\dfrac{e^{-a(r)s}-e^{-b(r)s}}{s}.$$
Apply Lemma~3 multiplying $n$ factors and dividing by $s$:
$$g(s)=\dfrac{(e^{-a(r)s}-e^{-b(r)s})^n}{s^{n+1}}=$$
$$=\sum\limits_{k=0}^n (-1)^k \binom{n}{k}\dfrac{e^{-a(r)s(n-k)-b(r)sk}}{s^{n+1}}.$$
After expanding the binom, compute the inverse Laplace transform of each term
 by Lemma~2(d) for the parameters $\al=0$, $\be=a(r)(n-k)+b(r)k$, $m=n$ as follows:
 $v_n(r,l)=$
 $$=\LT^{-1}\{g(s)\}=\sum\limits_{k=0}^n (-1)^k \binom{n}{k}\dfrac{(l-a(r)(n-k)-b(r)n)^n}{n!}.$$
To get the final formula for the conditional probability $P_n^c=v_n(R,L)/v_n(L,L)$ of
 connectivity it remains to substitute $a(L)=a(R)=a$, $b(L)=b$, $b(R)=R$ and cancel $n!$
\medskip

Now we prove the estimate for a minimal number of sensors
 making the network connected with a given probability $p$.
The condition $n\geq 1+(L-b)/a$, i.e. $L-a(n-1)-b\leq 0$, implies that the denominator
 of $P_n^c$ from the Constant Corollary is equal to $(L-an)^n$ corresponding to $m=0$.
Another assumption $(a+b)/2\leq R$ means that $b\leq 2R$, hence the numerator
 of $P_n^c$ contains only the first 2 terms, namely $(L-an)^n-n(L-a(n-1)-R)^n$.
The equivalent inequalities
 $$P_n^c=1-n\left(\dfrac{L-a(n-1)-R}{L-an}\right)^n\geq p,$$
 $$\left(\dfrac{L-an}{L-a(n-1)-R}\right)^n\geq \dfrac{n}{1-p},$$
 $$\left(1+\dfrac{R-a}{L-a(n-1)-R}\right)^n\geq \dfrac{n}{1-p}$$
 are weaker than $\left(1+\dfrac{R-a}{b-R}\right)^n\geq \dfrac{n}{1-p}$ as $L-a(n-1)\leq b$.
By Lemma~4 for $Q=\dfrac{b-R}{R-a}$ the last inequality holds if
 $$n\geq \dfrac{1}{2}\left(3(1-Q)+\sqrt{(3Q-1)^2+24Q^2\left(\dfrac{Q}{1-p}-1\right)}\;\right).$$
Since $(a+b)/2\leq R$ then $Q\leq 1$ and we replace $Q$ by 1
 in the last expression making the condition on $n$ only stronger:
$$n\geq \dfrac{1}{2}\left(3+\sqrt{4+24\left(\dfrac{1}{1-p}-1\right)}\right)
 =\dfrac{3}{2}+\sqrt{\dfrac{1+5p}{1-p}}.
\eqno{\square}$$
\smallskip

\noindent
\emph{Proof of the Exponential Corollary}. \\
By the Connectivity Theorem and Lemma~1 it suffices to compute $v_n(r,l)=(f^{[r]})^{(n*)}*u(l)$,
 where the probability density function is $f^{[r]}(l)=e^{-\la l}(u(l)-u(l-r))$ without extra factors.
Lemma~2(d) for $\al=\la$, $\be=r$ implies that
$$\LT\{e^{-\la(l-r)}u(l-r)\}(s)=\dfrac{e^{-rs}}{s+\la},$$
$$\LT\{f^{[r]}(l)\}(s)=\dfrac{ 1-e^{-r(s+\la)} }{s+\la}.$$
By Lemma~3 one has $v_n(r,l)=\LT^{-1}\{g(s)\}$, where
$$g(s)=\dfrac{(1-e^{-r(s+\la)})^n}{s (s+\la)^n}=
 \sum\limits_{i=0}^n(-1)^i\binom{n}{i}\dfrac{e^{-ir(s+\la)}}{s(s+\la)^n}.$$
The following result will be easily proved later.
\medskip

\noindent
{\bf Lemma~5.}
For any $\la>0$ and $n>0$ one has
$$\dfrac{1}{s(s+\la)^n}=\dfrac{1}{\la^n s}
  -\sum\limits_{j=0}^{n-1}\dfrac{1}{\la^{n-j}(s+\la)^{j+1}}.$$
\medskip

By Lemma~2(d) for $\al=\la$, $\be=ir$ one has
$$\LT^{-1} \left\{ \dfrac{e^{-irs} }{ (s+\la)^{j+1} } \right\}(l)=
 e^{-\la(l-ir)} \dfrac{ (l-ir)^{j} }{j!}u(l-ir).$$
It remains to apply Lemma~5, collect all terms in one sum and
 replace $u(l-ir)$ by $i<l/r$, i.e. $v_n(r,l)=\LT^{-1}\{g(s)\}=$
$$=\sum\limits_{i=0}^{i<l/r} \dfrac{(-1)^i}{\lambda^n}\binom{n}{i}
 \left(1- e^{-\lambda(l-ir)} \sum\limits_{j=0}^{n-1} \dfrac{\la^j(l-ir)^j}{j!} \right).
 \eqno{\square}$$
\medskip

\noindent
\emph{Proof of Lemma~5} is by induction on $n$.
The base $n=1$
$$\dfrac{1}{s(s+\la)}=\dfrac{1}{\la s}-\dfrac{1}{\la(s+\la)}\mbox{ is absolutely trivial}.$$
The induction step from $n-1$ to $n$ uses the base for $n=1$:
$$\dfrac{1}{s(s+\la)^n}=\left( \dfrac{1}{\la^{n-1} s}
 -\sum\limits_{j=0}^{n-2}\dfrac{1}{\la^{n-j-1}(s+\la)^{j}}\right)\dfrac{1}{s+\la}=$$
$$=\dfrac{1}{\la^n s}-\dfrac{1}{\la^n(s+\la)}
 -\sum\limits_{j=1}^{n-1}\dfrac{1}{\la^{n-j}(s+\la)^{j+1}}.
 \eqno{\square}$$
\smallskip

In the proof of the Normal Corollary we apply the following estimate
 for iterated convolutions of truncated probability densities
 using tails the normal density $f(s)$ over $\R$.
\medskip

\noindent
{\bf Lemma 6.}
Let $f(s)=\dfrac{1}{2\pi\sqrt{\si}}\exp\left(-\dfrac{(s-\mu)^2}{2\si^2}\right)$
 be \\ the normal density with a mean $\mu$ and deviation $\si$.
Then $$v_n(r,l)=(f^{[r]})^{(n*)}*u(l)\geq P(\sum_{i=1}^n y_i\leq l)-n\ep,\mbox{ where}$$
 $\ep=1-\int\limits_0^r f(s)ds$ and $y_i$ have the density $f(s)$ over $\R$.
\medskip

\noindent
\emph{Proof of Lemma 6} is by induction on $n$.
The base $n=1$:
$$v_1(r,l)=\int\limits_{-\infty}^{+\infty}f^{[r]}(s)u(l-s)ds
 =\int\limits_{-\infty}^{+\infty}f(s)u(l-s)ds-$$
$$-\int\limits_{\R-[0,r]}f(s)u(l-s)ds \geq \int\limits_{-\infty}^l f(s)ds-\ep=P(y_1\leq l)-\ep$$
$$\mbox{ since }u(l-s)\leq 1,\; \int\limits_{\R-[0,r]}f(s)u(l-s)ds\leq \int\limits_{\R-[0,r]}f(s)ds=\ep.$$
The induction step from $n-1$ to $n$ is similar:
$$v_n(r,l)=\int\limits_{-\infty}^{+\infty}f^{[r]}(s)v_{n-1}(r,l-s)ds=$$
$$\int\limits_{-\infty}^{+\infty}f^{[r]}(s)P(\sum_{i=1}^{n-1} y_i\leq l-s) ds
 -(n-1)\ep\int\limits_0^r f(s)ds \geq $$
$$\geq \int\limits_{-\infty}^{+\infty} f(s)P(\sum_{i=1}^{n-1} y_i\leq l-s)ds
 -\int\limits_{\R-[0,r]}f(s)ds-(n-1)\ep$$
$$\geq P(\sum_{i=1}^n y_i\leq l)-n\ep \mbox{ using } P(\sum_{i=1}^{n-1} y_i\leq l-s)\leq 1.
\eqno{\square}$$

\noindent
\emph{Proof of the Normal Corollary}. \\
By the Connectivity Theorem the probability of connectivity is
 $P_n=v_n(R,L)/v_n(L,L)$, where the denominator
 $v_n(L,L)=(f^{[L]})^{(n*)}*u(L)$ is computed using
 the truncated normal density over $[0,L]$, while
 in $v_n(R,L)=(f^{[R]})^{(n*)}*u(L)$
 the same density is truncated over the smaller range $[0,R]$.
\smallskip

As usual we may forget about extra constants in front of
 $f(s)=\dfrac{1}{\si\sqrt{2\pi}}\exp\left(-\dfrac{(s-\mu)^2}{2\si^2}\right)$.
For a given probability $p$ we will find a condition on $n$ such that $P_n\geq p$.
We will make the inequality $P_n\geq p$ simpler and stronger replacing
 $v_n(L,L)$ and $v_n(R,L)$ by their upper and lower bounds, respectively.
\smallskip

The denominator $v_n(L,L)$ is the iterated convolution of normal densities truncated over $[0,L]$.
This convolution of positive functions becomes greater if we integrate the same functions over $\R$.
Then $v_n(L,L)\leq P(\sum_{i=1}^n y_i\leq L)$,  probability that the sum of
 $n$ normal variables with the mean $\mu$ and deviation $\si$ is not greater than $L$.
The sum $\sum_{i=1}^n y_i$ is the normal variable with the mean $n\mu$ and deviation $\si\sqrt{n}$.
\smallskip

Then $P(\sum_{i=1}^n y_i\leq L)=\Phi\left(\dfrac{L-n\mu}{\si\sqrt{n}}\right)$,
 where the standard normal distribution is
 $\Phi(x)=\dfrac{1}{\sqrt{2\pi}}\int\limits_{-\infty}^x e^{-s^2/2}ds$.
Taking into account the lower estimate of $v_n(R,L)$ from Lemma 6,
 we replace the inequality $P_n\geq p$ by the stronger one
 $$1-n\ep/P(\sum_{i=1}^n y_i\leq L)\geq p \mbox{ or } \Phi\left(\dfrac{L-n\mu}{\si\sqrt{n}}\right)\geq\dfrac{n\ep}{1-p}.$$
Split the last inequality into two simpler ones:
$$\Phi\left(\dfrac{L-n\mu}{\si\sqrt{n}}\right)\geq p \mbox{ and }p\geq\dfrac{n\ep}{1-p}.$$
The latter inequality gives $n\leq p(1-p)/\ep$ as expected, where
$$\ep=\dfrac{1}{\si\sqrt{2\pi}}\int\limits_{\R-[0,R]} f(s)ds=
 \Phi\left(-\dfrac{\mu}{\si}\right)+1-\Phi\left(\dfrac{R-\mu}{\si}\right).$$
The former inequality above becomes the quadratic one:
$$L-n\mu\geq\si\Phi^{-1}(p)\sqrt{n}, \quad \mu n+\si\Phi^{-1}(p)\sqrt{n}-L\leq 0.$$
The final condition says that $n$ is not greater than the square of the 2nd root
 $(\sqrt{4\mu L+\si^2\Phi^{-2}(p)}-\si\Phi^{-1}(p))/2\mu$.
\hfill
$\square$
\medskip


\section{Networks with sensors of different types}

We derive an explicit formula and algorithm for
 computing the probability of connectivity when
 distances between successive sensors have different constant densities.
These general settings might be helpful for \emph{heterogeneous} networks
 containing sensors of different types, e.g. of different transmission radii.
Assume that each distance between successive sensors has one of $k$ constant densities
 $f_j(l)=c_j$ on $[a_j,b_j]\subset[0,L]$ and $f_j(l)=0$ otherwise, $j=1,\dots,k$.
The condition $\int\limits_0^{L} f_j(l)dl=1$ implies that $1/c_j=b_j-a_j$.
\smallskip

Note that the types of densities may not respect the order of sensors in $[0,L]$,
 e.g. the 1st and 3rd distances can be from the 2nd group of densities equal to $f_2(l)$,
 while the 2nd distance can be from the 1st group.
In this case we say that index~1 belongs to group 2, symbolically
$(1)=2$. Here the brackets $(\cdot)$ denote the operator transforming
 an index $i=1,\dots,n$ of a distance into its group number $(i)$
 varying from 1 to $k$.
\smallskip

For a heterogeneous network, the function $v_n(r,l)$ from section~III
 will be a sum over arrays of signs $Q=(q_1,\dots,q_n)$
 depending on prescribed densities $\{f_1,\dots,f_k\}$.
Let $[a_j(r),b_j(r)]$ be the intersection of $[0,r]$
 with $[a_j,b_j]$, where $f_j\neq 0$.
Set $q_i^{\pm}=(1\pm q_i)/2$, e.g. $1^+=1$, $1^-=0$.
\medskip

\noindent
{\bf The Heterogeneous Corollary.}
In the above notations and under the conditions of the Connectivity Theorem
 assume that distances between successive sensors have probability densities
 $f_j(l)=c_j$ on $[a_j,b_j]$, $j=1,\dots,k$.
Then the probability of connectivity is $P_n=\dfrac{v_n(R,L)}{v_n(L,L)}$, where
$$v_n(r,l)=\sum\limits_{Q=(q_1,\dots,q_n)}^{\ab{Q}<l}
 \dfrac{d_Q}{n!}(l-\ab{Q})^n \mbox{ and}$$
$$d_Q=\prod\limits_{i=1}^n (-1)^{q_i^+}c_{(i)},\quad
 \ab{Q}=\sum\limits_{i=1}^n ( a_{(i)}(r)q_i^- + b_{(i)}(r) q_i^+ ).$$
\smallskip

The indices in the brackets $(i)$ from the last formula above
 take values $1,\dots,k$ for each $i=1,\dots,n$, i.e.
 $[a_{(i)}(r),b_{(i)}(r)]$ is the segment where the $(i)$-th density $f_{(i)}$
 is defined after restricting it to the transmission range $[0,r]$.
In particular, if each $i$-th distance has its own density then
 $(i)=i$ and the indices $i,j=1,\dots,n$ are equal to each other.
\smallskip

First we show that the Constant Corollary is a very partial case of the Heterogeneous Corollary
 with only one constant density $f_1=1/(b-a)$ on $[a,b]$, i.e. $k=1$.
To compute $v_n(L,L)$ we note that $[a_1(L),b_1(L)]=[a,b]$.
Let $k$ be the number of pluses in an array $Q$.
Then $d_Q=(-1)^k/(b-a)$ and $\ab{Q}=a(n-k)+bk$.
So the sum over $Q$ can be rewritten as a sum over $0\leq k\leq n$.
For any fixed $k$ there are $\binom{n}{k}$ different
 arrays $Q$ containing exactly $k$ pluses.
By the Heterogeneous Corollary the common term in the sum $v_n(L,L)$ over $k$ is
 $(-1)^k  \binom{n}{k} (L-a(n-k)-bk)^n$.
The only difference in computing $v_n(R,L)$ is that $b_1(R)=R$,
 which leads to the formula from the Constant Corollary.
\smallskip

The complexity to compute the function $v_n(r,l)$ from the Heterogeneous Corollary
 is $O(2^n)$, because $v_n(r,l)$ is a sum over $2^n$ arrays of signs and
 $\ab{Q}$ is a weighted sum of endpoints $a_i(r),b_i(r)$.
In the general case, the expression $\ab{Q}$ can take $2^n$ different values.
If there are only $k$ different endpoints then
 the algorithm has the polynomial complexity $O(n^k)$,
 see the 3-step Density Corollary in Appendix~D.
If all the segments $[a_j,b_j]$ are subsets of $[0,R]$
 then any network will be connected and the formula above
 gives 1, because the numerator of $P_n$ coincides with the denominator
 when $a_j(R)=a_j(L)$ and $b_j(R)=b_j(L)$, $j=1,\dots,k$.
\medskip

\noindent
\emph{Proof of the Heterogeneous Corollary}
 extends the proof of the Constant Corollary.
We consider the truncated densities
$$f_i^{[r]}(l)=c_{(i)}(u(l-a_{(i)}(r))-u(l-b_{(i)}(r))),$$
 $i=1,\dots,n$, where $(i)$ denotes the group containing the $i$th distance.
By Lemma~2(c) for $m=0$ one has
$$\LT\{ f_i^{[r]} \}(s)=c_{(i)}\dfrac{ e^{-a_{(i)}(r)s} - e^{-b_{(i)}(r)s} }{s}.$$
Substitute each Laplace transform $\LT\{ f_i^{[r]} \}(s)$ into the function $g(s)$
 from Lemma~3 and expand the product $g(s)$, which gives the following sum of $2^n$ terms:
$$g(s)=\dfrac{1}{s}\prod\limits_{i=1}^n c_{(i)} \dfrac{ e^{-a_{(i)}(r)s} - e^{-b_{(i)}(r)s} }{s}
 =\sum\limits_{Q} d_Q \dfrac{ e^{-\ab{Q}s} }{s^{n+1}}.$$
The sum is taken over arrays $Q=(q_1,\dots,q_n)$ of signs.
The sign $q_i=-1$ means that the term with $a_{(i)}(r)$ is taken from the $i$-th factor,
 the sign $q_i=+1$ encodes the second term with $b_{(i)}(r)$.
The total power of the exponent in the resulting term corresponding to $Q$
 is $-\ab{Q}s$, where $\ab{Q}=\sum\limits_{i=1}^n ( a_{(i)}(r)q_i^- + b_{(i)}(r) q_i^+ )$.
So each minus contributes $-a_{(i)}(r)s$ to the total power,
 while each plus contributes $-b_{(i)}(r)s$.
Each plus contributes factor $(-1)$ to the coefficient $d_Q$,
 i.e. $d_Q=\prod\limits_{i=1}^n (-1)^{q_i^+}c_{(i)}$ as required.
\medskip

Compute the inverse Laplace transform by Lemma~2(d):
$$v_n(r,l)=\LT^{-1}\{g(s)\}=\sum\limits_{Q} \dfrac{d_Q}{n!}  (l-\ab{Q})^n u(l-\ab{Q}),$$
 where the unit step functions $u(l-\ab{Q})$ can be replaced
 by the upper bound $l<\ab{Q}$ as in the final formula.
\hfill $\square$
\medskip

The algorithm for computing the function $v_n(r,l)$
 from the Heterogeneous Corollary has the following steps:
\smallskip

\noindent
$\bu$
initialise 2 arrays $a_{(i)}(r),b_{(i)}(r)$, $i=1,\dots,n$;
\smallskip

\noindent
$\bu$
make a computational loop over $2^n$ arrays $Q$ of signs;
\smallskip

\noindent
$\bu$
for each $Q$ compute $\ab{Q}$ and check the upper bound $l\leq\ab{Q}$,
find $d_Q(l-\ab{Q})^n$ and add it to the current value of $v_n(r,l)$.
\medskip

The algorithm for computing $v_n(L,L)$ is similar, replace $R$ by $L$.
If we are interested only in $P_n$, we may forget about $n!$
 which is canceled after dividing $v_n(R,L)$ by $v_n(L,L)$.
\smallskip




\section{Piecewise Constant Densities}

In this appendix we show how to compute the probability of connectivity building
 any piecewise constant density from elementary blocks in the Heterogeneous Corollary.
The building engine is the Average Density Corollary below dealing with
 the average $f(s)=\sum\limits_{j=1}^k f_j(s)/k$ of constant densities
 $f_j(s)=c_j$ on $[a_j,b_j]$ and $f_j(s)=0$ otherwise, $j=1,\dots,k$.
The factor $1/k$ guarantees the condition $\int_0^L f(s)ds=1$,
 which follows from $\int_0^L f_j(s)ds=1$, $j=1,\dots,k$.
\smallskip

For any ordered partition $n=n_1+\dots+n_k$ into $k$ non-negative integers,
 denote by $(n_1,\dots,n_k)$ the collection of densities, where
 the first $n_1$ densities equal $f_1$, the next $n_2$ densities equal $f_2$ etc.
For example, given 2 constant densities $f_1,f_2$, number $n=3$ can be split
 into 2 non-negative integers in one of the 4 ways:  $3=0+3=1+2=2+1=3+0$.
Then $(1,2)$ denotes the collection $(f_1,f_2,f_2)$, i.e. the 1st distance in such a network
 has the density $f_1$, while the remaining 2 distances have the density $f_2$.
For each partition $(n_1,\dots,n_k)$ or, equivalently, a collection of constant densities,
 let $v_n^{(n_1,\dots,n_k)}(r,l)$ be the function defined by the formula
 from the Heterogeneous Corollary in Appendix C.
\medskip

\noindent
{\bf The Average Density Corollary.}
In the above notations and under the conditions of the Connectivity Theorem if
 distances between successive sensors have the probability density
 $f(l)=\sum\limits_{j=1}^k f_j(l)/k$ on $[0,L]$,
 then the probability of connectivity is
 $P_n=\dfrac{\sum v_n^{(n_1,\dots,n_k)}(R,L)/n_1!\dots n_k! }{
 \sum v_n^{(n_1,\dots,n_k)}(L,L)/n_1!\dots n_k! }$.
Both sums are taken over all collections of densities ${(n_1,\dots,n_k)}$
 corresponding to ordered partitions $n=n_1+\dots+n_k$.
\medskip

The products $n_1!\dots n_k!$ can not be canceled in the formula
above,
 because the numerator and denominator of $P_n$ are sums of
 many terms involving different products $n_1!\dots n_k!$
 over all ordered partitions $n=n_1+\dots+n_k$.
The complexity to compute $P_n$ is $O(n 2^n)$, because
 each function $v_n^{(n_1,\dots,n_k)}$ is computed by the algorithm
 describe after the Heterogeneous Corollary using $O(2^n)$ operations.
In partial cases the computational complexity can be reduced to
 polynomial, see comments after the 3-step Density Corollary below.
The algorithm computing the probability from the Average Density Corollary
 applies the algorithm from the Heterogeneous Corollary to each function
 $v_n^{(n_1,\dots,n_k)}(R,L)$ and $v_n^{(n_1,\dots,n_k)}(L,L)$
 substituting the results into the final formula above.
\medskip

\noindent
\emph{Proof of the Average Density Corollary.}\\
We may forget about the factor $1/k$ as usual.
Set $g_j(s)=\LT\{f_j^{[r]}\}$, $j=1,\dots,k$.
Lemma~3 implies that $v_n(r,l)=\LT^{-1}\{g(s)\}$,
 where $g(s)=(\sum\limits_{j=1}^k g_j(s))^n/s$.
Expand the brackets:
 $g(s)=\sum\limits \dfrac{n!}{n_1!\dots n_k!} \dfrac{g_1^{n_1} \dots g_k^{n_k}}{s}$,
 where the sum is over all partitions $n=n_1+\dots+n_k$ into $k$ non-negative integers.
\smallskip

By Lemma~3 each term $g_1^{n_1} \dots g_k^{n_k}/s$ is
 the inverse Laplace transform of the function $v_n^{(n_1,\dots,n_k)}(r,l)$,
 where the first $n_1$ distributions are $f_1$,
 the next $n_2$ distributions are $f_2$ etc.
It remains to cancel $n!$ in the final expression.
\hfill $\square$
\medskip

By taking sums of constant densities $c_j$ on $[a_j,b_j]$,
 one can get any piecewise constant function on $[0,L]$.
Any reasonable function can be approximated by piecewise constant ones.
Hence the Heterogeneous Corollary and Average Density Corollary are building blocks
 for computing the probability of connectivity for any real-life deployment of sensors.
\smallskip

We demonstrate this universal approach for the sum of
 2 constant densities over 2 different segments.
So the density in question is a 3-step function  depending on the radius $R$
 and one more parameter $C$, its graph is shown in Fig.~9.
Let $f=(f_1+f_2)/2$ be the density on $[0,L]$ such that
$$f_1(l)/2=\left\{ \begin{array}{ll}
C & \mbox{ if } l\in[0,R],\\
0 & \mbox{ otherwise; }
 \end{array} \right.$$
$$f_2(l)/2=\left\{ \begin{array}{ll}
1/R-C & \mbox{ if } l\in[R/2,3R/2],\\
0 & \mbox{ otherwise. }
 \end{array} \right.$$
 where $0<C<1/R$ is a constant and $3R/2\leq L$, see Fig.~9.
The constants $C$ and $1/R$ are chosen so that $\int\limits_0^L f(l)dl=1$.
\smallskip

\begin{figure}[!h]
\centering
\includegraphics[width=2.3in]{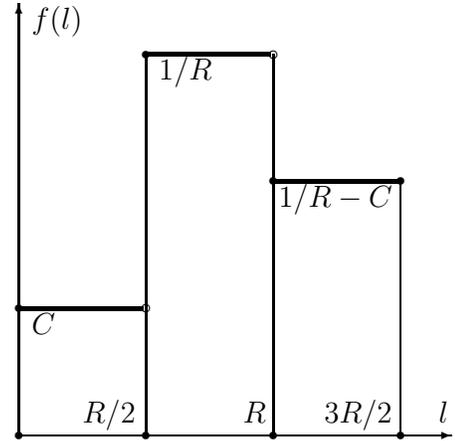}
\caption{The piecewise constant distribution depending on $R,C$}
\label{Fig9}
\end{figure}
\smallskip

From Fig.~9 for a network of a sink node at 0 and 1 sensor at $y_1$
 the probability of connectivity is $P(0\leq y_1\leq R)=(CR+1)/2$,
 the area of the first two rectangles below the graph of $f(l)$.
For example, if $C=0.9/R$ then $P_1=0.95$ as shown in Fig.~10, so
 it is very likely that 1 sensor will be close enough to the sink,
 although such a network can not cover the whole segment $[0,L]$.
The 3-step Density Corollary below gives an example how to compute
 the probability of connectivity explicitly for a piecewise constant density
 using the Heterogeneous Corollary and Average Density Corollary.
\medskip

\noindent
{\bf The 3-step Density Corollary.}
Under the conditions of the Connectivity Theorem and
 for the piecewise constant density $f(l)$ above,
 the probability of connectivity is $P_n=$
$$\dfrac{ \sum\limits_{m=0}^n \sum\limits_{k_1=0}^m \sum\limits_{k_2=0}^{n-m}
   \dfrac{(-1)^{k_1+k_2} (L-(2k_1+k_2+n-m)R/2)^n}{d_m k_1!(m-k_1)!k_2!(n-m-k_2)!} }{
  \sum\limits_{m=0}^n \sum\limits_{k_1=0}^m \sum\limits_{k_2=0}^{n-m}
   \dfrac{(-1)^{k_1+k_2} (L-(2k_1+2k_2+n-m)R/2)^n}{d_m k_1!(m-k_1)!k_2!(n-m-k_2)!} },$$
 where $d_m=C^{-m}(1/R-C)^{m-n}$, the sums are over all possible values of $m,k_1,k_2$
 such that the expressions in the brackets taken to the power $n$ are positive.
\medskip

The complexity to compute the probability $P_n$ above is $O(n^3)$,
 because the sums in the numerator and denominator are over
 3 non-negative integers not greater than $n$
 and each term requires $O(1)$ operations.
If $C=1/R$, i.e all distances are in $[0,R]$, then set $d_m=0$ for $m<n$.
Hence $m=n$, $k_2=0$ and the sums over 3 parameters $m,k_1,k_2$ reduce to the same single sum
 over $k_1=0,\dots,n$ in the numerator and denominator, which gives $P_n=1$ as expected for $C=1/R$.
\smallskip

If $C=0$, i.e. each distance is uniformly distributed
 on $[R/2,3R/2]$, then set $d_m=0$ for $m>0$.
Therefore, $m=0$, $k_1=0$ and the result containing only sums over $k_2=0,\dots,n$
 coincides with the probability $P_n^c$ from the Constant Corollary with $[a,b]=[R/2,3R/2]$ after
 canceling $d_0$ and multiplying the numerator and denominator by $k_2!$ to get $\binom{n}{k_2}$
$$P_n=\dfrac{ \sum\limits_{k_2=0}^n (-1)^{k_2} (L-(k_2+n)R/2)^n/k_2!(n-k_2)! }{
  \sum\limits_{k_2=0}^n (-1)^{k_2} (L-(2k_2+n)R/2)^n/k_2!(n-k_2)! }.$$
\smallskip

In the 3-step Density Corollary for $n=1$ both sums contain
 only 4 non-zero terms corresponding to the parameters
$$(m,k_1,k_2)=(0,0,0);\; (0,0,1);\ (1,0,0);\ (0,1,0).$$
Then all the factorials in the formula are 1 and we get $P_1=$
$$\dfrac{ (\frac{1}{R}-C)(L-\frac{R}{2})-(\frac{1}{R}-C)(L-R)+CL-C(L-R) }{
 (\frac{1}{R}-C)(L-\frac{R}{2})-(\frac{1}{R}-C)(L-\frac{3R}{2})+CL-C(L-R) }$$
 $=(CR+1)/2$ as we have checked using Fig.~9 directly.
\medskip

\noindent
\emph{Proof of the 3-step Density Corollary.}\\
In the notations of the Heterogeneous Corollary we have only $k=2$ densities.
Let the first $m$ distances between successive sensors have the probability density $f_1$,
 while the last $n-m$ distances have the density $f_2$.
An array $Q$ of $n$ signs similarly splits into two parts
 consisting of $m$ signs and $n-m$ signs.
Let $k_1$ and $k_2$ be the number of pluses in each part.
\medskip

To compute $v_n^{(m,n-m)}(L,L)$ from
 the Heterogeneous Corollary for the partition $n=m+(n-m)$ we note that
$$[a_1(L),b_1(L)]=[0,R],\; [a_2(L),b_2(L)]=[R/2,3R/2],$$
$$d_Q=(-1)^{k_1+k_2}/d_m,\quad \ab{Q}=(k_1+k_2+(n-m)/2)R.$$
So the sum over arrays $Q$ can be rewritten as a sum over $k_1,k_2$.
For fixed values of these parameters, there are
 $\binom{m}{k_1}\binom{n-m}{k_2}$ different arrays of signs.
After canceling the factorials $m!$ and $(n-m)!$
 in the Average Density Corollary,
 the sum $\sum\limits_{m=0}^n v_n^{(m,n-m)}(L,L)$
 is the required denominator of $P_n$.
\medskip

The only difference in computing $v_n(R,L)$ is
 that $b_2(R)=R$, not $3R/2$.
This replaces $2k_2$ by $k_2$ in the numerator.
\hfill $\square$
\smallskip

\begin{figure}[!h]
\centering
\includegraphics[width=3.5in]{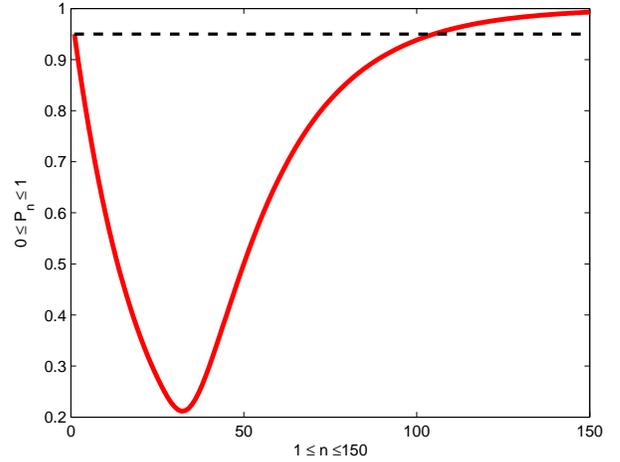}
\caption{The probability of connectivity for
 $f(l)$ with $C=0.9/R$}
\label{Fig10}
\end{figure}
\smallskip

Given the piecewise constant density $f(l)$ with
 the intermediate parameter $C=0.9/R$,
 Table~6 shows the minimal number of sensors having
 different radii such that the network in $[0,L]$ is
 connected with probability 0.95, where $L=1$km.
Fig.~10 shows the probability of connectivity $P_n\geq 0.2$.
\medskip

\noindent
{\bf Table 6.} The probability of connectivity for
 the piecewise constant density with $C=0.9/R$ and different radii.
\bigskip

\hspace*{-5mm}
\begin{tabular}{|l|c|c|c|c|c|}
\hline
Transmission Radius, m. & 250 & 200 & 150 & 100 & 50\\
\hline

Min Number of Sensors & 12 & 17 & 25 & 44 & 105\\
\hline
\end{tabular}
\medskip


\section*{Acknowledgment}

\noindent
We acknowledge the support of the UK MOD Data and
 Information Fusion Defence Technology Centre,
 project DTC.375 `Ad hoc networks for decision making and object tracking'.
\ifCLASSOPTIONcaptionsoff
  \newpage
\fi

\bibliographystyle{IEEEtran}

\end{document}